\documentclass[12pt,oneside,english]{article}
 \usepackage{epsfig}

\usepackage[T1]{fontenc}
\usepackage[latin1]{inputenc}
\usepackage{babel}
 \makeatletter
 
\begin{document}
 \bigskip \noindent 
\centerline{\bf \large Ordered and periodic chaos of the bounded one-dimensional
multibarrier potential }

\bigskip

\centerline{\large  D. Bar$^{a}$} 

\bigskip 

\centerline{$^a$Department of Physics, Bar Ilan University, Ramat Gan,
Israel  }

\bigskip \bigskip

\begin{abstract}

\noindent

{\it Numerical analysis indicates that there exists an unexpected new ordered
chaos for the bounded one-dimensional multibarrier potential.   
For certain values of  
the number of barriers,  repeated identical forms (periods) 
 of  the wavepackets result upon  passing through
the multibarrier potential.  }
\end{abstract}

\bigskip \bigskip


\bigskip \bigskip

{{\bf  Pacs}:  05.45.Pq, 02.10.Yn, 03.65.Nk}


\bigskip \bigskip

\pagestyle{myheadings}
\markright{INTRODUCTION}
\bigskip \noindent 
\protect \section{Introduction \label{sec1}}

Recent studies of the chaotic aspects of different complex systems have 
resulted in finding certain conditions 
under which  the related chaos become {\it ordered}, predicted 
 and deterministic.  
Among these  one may  
include the early ordered chaos found in the one-dimensional logistic maps
\cite{Kaneko} or the optical ``chaos itinerancy'' \cite{Arecchi} in which the 
onset of chaos  is done  in a rather ordered manner. 
 Another
known example of ordered chaos has been shown in \cite{Hondou} for the
one-dimensional periodic potential with constant slope in which the time series
are produced by a tent map \cite{Hondou}. The particle which passes through such
a potential is shown \cite{Hondou} to be under the influence of 
 an ordered chaos which is
effected in an unexpected drift  in the opposite direction to that of the
average potential gradient \cite{Hondou}. It is also shown  \cite{Hondou} for
the case of zero  average gradient that the smaller 
 becomes the widths of the potential barriers  
 the more increased  is the transition probability and the 
 corresponding ordered chaos
\cite{Hondou}.  \par 
We  note   that similar results were  shown \cite{Bar1}  regarding the bounded 
one-dimensional 
multibarrier potential  which  certainly has no average gradient and in which 
the width of
its barriers is inversely proportional to the number of them. It is found
\cite{Bar1}, analogously to \cite{Hondou},   
  that the larger is the number of barriers along the
same spatial length,  which means that the smaller are their width, 
the higher  becomes the 
transition probability of this system.  Thus, since the bounded multibarrier 
system was shown in
\cite{Bar1} to be chaotic then, according to the criterion in \cite{Hondou}
 for ordered chaos,   we conclude that this system  
 belongs  to the last class. \par 
 We note that although the possible presence of 
quantum chaos is generally expected to exist in systems in which the classical 
limit is chaotic \cite{Lewenkopf}, there are, nevertheless, quantum systems which 
show  chaotic signs without classical counterparts. This has been  
shown in \cite{Ashkenazy}  for the quantum one-dimensional single  barrier 
 and in \cite{Bar1} for the bounded one-dimensional multibarrier potential. Also, the appearance 
of chaotic behaviour in these one-dimensional systems in \cite{Bar1,Ashkenazy} 
is due to their being bounded and composed of a not large number of barriers.  
Thus,  in the limit of very large number 
of barriers arrayed along the whole axis as,  for example, the one-dimensional 
Kronig-Penney system  \cite{Merzbacher}, one should not expect chaotic effects (see 
also the discussion in \cite{Ingold} of the localization-delocalization 
transition).   
 \par
 Our main goal in this work  is to show  that changing the number or 
 (and) the width 
 of the barriers 
results in the appearance of {\it  new} ordered chaos which is effected in the
form of periods.  These are neither  periods in time nor in space but  periods in the
number of barriers $N$.  We note that by the phrase new  ordered chaos we do 
not mean that 
the observed wavepackets become less chaotic and complex but that the same chaotic 
structure is seems to be repeatedly observed as the number of 
barriers $N$ increases 
by  specific values.      \par
 We note, as mentioned,   that 
chaoticlike effects were discussed in \cite{Ashkenazy} with respect to {\it one}
 rectangular barrier
and it was found that the chaotic appearance of the passing wavepacket as well
as its correlation with the initial one critically depend upon the width of the
single barrier.   
Thus, it seems appropriate to extend the discussion to the multibarrier potential 
 and find the conditions under which  the related  chaotic effects  become 
periodically  ordered.  
   \par  
We use in our discussion the  well known fact that a classical  
(or semiclassical) wavepacket  
spreads and becomes chaotic 
\cite{Bar1,Ashkenazy,Zaslavsky,Schieve,Reichl} when it pass through a 
region along which a system of potential
barriers (or wells) is arranged.  The degree of the resulting chaos may be 
determined from the 
correlation \cite{Havlin} between the initial and final forms of 
the passing particle 
(wavepacket).  Thus, if this correlation turns out to be small then the
initial wavepacket has been considerably changed and its chaotic effects have
increased in the passage through the
barriers.  In the following we use the Lanczos tridiagonalization method
\cite{Lanczos,Cullum,Maple} for
calculating the correlation (and, therefore, the degree of chaos)  between the initial 
wavepacket and the  one which
emerges from the potential array.      \par
We show,  using the obtained data of the correlation,  
  that  the passing chaotic wavepackets are, under cetain
values of $N$ and $c$ (which is the ratio of total interval to total width of
the potential array), strictly periodic and predicted. That is, 
 suppose that a specific wavepacket which  pass through a multibarrier
potential with a given $N$,  $c$ and total length $L$   
assumes some specific form.  Then it is shown that the same initial wavepacket 
 assumes  exactly the same    form  with the corresponding 
correlation when  it pass through $(N+nP)$ barriers arranged along the same 
length $L$ and $c$ where $n$ is the positive numbers $1, 2, 3, ...$  and 
$P$ are  the periods.  \par
We also show that although the passing wavepackets are highly sensitive to
variations of the ratio $c$, there are, nevertheless, some specific values of
$N$ for which the waveforms and the corresponding correlations are preserved 
for whole ranges
of $c$.  \par
 In Section 2 we present the bounded one-dimensional multibarrier potential and
 the formalism of the energy level statistics \cite{Bar1,Reichl} used for
 discussing its chaotic properties. This method were used in \cite{Bar1} for 
 demonstrating that this system is chaotic. We use here numerical analysis for
 further  discussing the mentioned periodic ordered aspect of  these 
 chaotic effects. 
 We  show the
  dependence of  the passing wavepackets (and their corresponding 
 correlations with the given initial one)  upon the number of barriers $N$ and the
  ratio $c$.  In Section 3 we demonstrate the mentioned  order 
  which is
  effected:  (1) through the remarked  periods $P$ which turns out to be  
   of two kinds;  
  one is very
frequent and  is effective for a large specific values of $N$ 
and the other is
rare and shows up only in two specific values of $N$ (from the range $2 \le N \le
72$)   and (2) through 
  the constancy of the passing
  wavepackets and their related correlations for certain ranges of the ratio
  $c$. In Table 1 we show  the correlations for some specific
  values of $N$ and for 6 different values of the ratio $c$ and indicate the
  correlations which are periodic.  We conclude in Section 4 with a summary of 
  the main points.

\pagestyle{myheadings}
\markright{THE CORRELATION BETWEEN THE INITIAL AND.......}
\bigskip \noindent 
\protect \section{The correlation between the initial and final 
wavepackets for the bounded
one-dimensional multibarrier potential \label{sec2}}

The bounded one-dimensional multibarrier potential discussed here is 
supposed to be arranged
along the $x$ axis between the points $x=-10$ and $x=10$. Assuming that the
number of barriers in the system is $N$ one may introduce \cite{Bar1} the
variables $a$ and $b$ which respectively denote the total width of the $N$
barriers, where the potential $V$ satisfies $V > 0$, and the total interval
among them where $V=0$. Thus, one may realize \cite{Bar1} that the width of each
barrier is $\frac{a}{N}$ and the interval between any two neighbouring ones is
$\frac{b}{(N-1)}$.  Denoting the ratio of $b$ to $a$ by $c$ and the total length
of the system $a+b$ by $L$ one may express \cite{Bar1} $a$ and $b$ in terms of $c$
and $L$ as 

\begin{equation} \label{e1}   a=\frac{L}{1+c} \ \ \ \ \ \ b=\frac{Lc}{1+c} 
\end{equation} 
The possible existence of chaotic properties for any bounded one-dimensional 
multibarrier (or multiwell)
potential system is usually determined by  applying the energy level
statistics \cite{Bar1,Reichl}. In this method one begins from the following
two-dimensional matrix equation 
\begin{equation}  \label{e2} \left[
\begin{array}{c} A_{2N+1} \\ B_{0} 
\end{array} \right]=\left[ \begin{array}{c c} S_{11}&S_{12} \\ S_{21} &S_{22} 
\end{array} \right]\left[ \begin{array}{c} A_{0} \\ B_{2N+1} 
\end{array} \right],  \end{equation}   
where $A_{2N+1}$ and $B_{2N+1}$ are the amplitudes of the transmitted and
reflected parts respectively of the passing wavapacket from  the $N$-th
potential barrier. $A_0$ is the transmission coefficient of the initial wave
that approach the first barrier and $B_0$ is the reflected part from this
barrier. The components $S_{11}$,  $S_{12}$, $S_{21}$, and  $S_{22}$ are the
matrix elements of the two-dimensional $S$ matrix which are related to the
corresponding transfer matrix $Q$ of the multibarrier  potential (see, for
example,  Eqs (21) in
\cite{Bar1}). The energy level statistics method \cite{Reichl} is used by
imposing boundary value conditions at the remote boundaries of the system. In
\cite{Bar1} periodic boundary conditions are used at the points $|x|=R$, where
$R$ is much larger than the total length  $L=a+b$ of the system, so that 
 one obtains 
\begin{eqnarray}  && A_{2N+1}f(R)=A_0f(-R)  \label{e3} \\  && 
B_{2N+1}f(-R)=B_0f(R),   \nonumber \end{eqnarray} 
where $f(R)$ and $f(-R)$ denote  the wavepackets at the points  $x=R$ and $x=-R$
respectively. Using Eqs (\ref{e3}) one may write Eq (\ref{e2}) as 
\begin{equation}  \label{e4} \left[
\begin{array}{c} A_{2N+1} \\ B_{0} 
\end{array} \right]=\frac{f(R)}{f(-R)} \left[ \begin{array}{c c} S_{11}&S_{12} \\ S_{21} &S_{22} 
\end{array} \right]\left[ \begin{array}{c} A_{2N+1} \\ B_{0} \end{array}  
\right]
 \end{equation}  
In order to obtain a nontrivial solution for the vector $\left[
\begin{array}{c} A_{2N+1} \\ B_{0} 
\end{array} \right]$  we must solve the following equation   
\begin{equation} \label{e5}  \det \left[ \begin{array}{c c} 
\frac{f(R)}{f(-R)}S_{11}-1&  \frac{f(R)}{f(-R)}S_{12}  \\
 \frac{f(R)}{f(-R)}S_{21} & \frac{f(R)}{f(-R)}S_{22}-1
\end{array} \right] =0 \end{equation} 
In \cite{Bar1} we have used for $f(R)  \ \ (f(-R))$ the plane wave $e^{ikR}  \ \
(e^{-ikR})$ and have
expressed the $S$ matrix elements in terms of the known transfer matrix elements
(see Eqs (15) and (21) in \cite{Bar1}).  As a result of these 
 substitutions one obtains from Eq (\ref{e5}), as in  \cite{Bar1},  a complex
 equation from which the appropriate energies which  correspond to its real and
 imaginary parts are derived. Figure (8) in \cite{Bar1} shows the level spacing
 distribution of these energies in the form of a histogram which is clearly of
 the chaotic Wigner type \cite{Reichl}. \par
 In this work we use, instead of plane waves, a semiclassical complex Gaussian 
 wavepacket since this kind of wave function tends easily to be deformed and
 becomes chaotic upon passing  a multibarrier (or multiwell) potential 
 \cite{Bar1}. Also, the semiclassical character of the wavepacket enables one 
 to simultaneously discuss, as  done in the following,  its momentum and 
 position. Note that even in the quantum regime one may introduce the coherent
 state formalism \cite{Glauber,Swanson} which allows one \cite{Swanson} to simultaneously
 define the expectation values of the conjugate variables $Q$ and $P$. We note 
 that in  the numerical part of this work all the  
  wavepackets (denoted $\phi$), including the initial one,   are numerically 
  and graphically 
  constructed from a given  complex Gaussian packet  $P_{packet}$ 
  given by  
 \begin{equation}
  \label{e6}
  P_{packet}(x,t,x_0,p_0,w_0)=\frac{\sqrt{w_0}\pi^{\frac{1}{4}}e^{-\frac{p_0^2}{4w_0^2}}
 e^{\frac{w_0^2(i(x_0-x)-\frac{p_0}{2w_0^2})^2}{1-2itw_0^2}}}{\sqrt{1-2itw_0^2}},
\end{equation}
where $x_0$ is the initial value of the mean position of the packet in
coordinate space and $p_0$ and $w_0$ are the initial momentum and width in $p$
space. The width $w_0$ is,  actually, the initial uncertainty in the momentum.
For an effective numerical simulation the space and time variables were
discretized \cite{Bar1,Maple} with a resolution of $dx=\frac{1}{7}$ and
$dt=\frac{1}{50}$ so that we  obtain $dx^2>dt$ which is necessary for stabilizing and
steadying the relevant numerical method \cite{Bar1,Maple}. For the initial
$x_0$, $p_0$ and $w_0$ we choose the values of $x_0=-10$, $p_0=3$ and
$w_0=\frac{1}{2}$. In the semiclassical discussion adopted here we assume that
the wavepacket is  associated with a particle of mass $m$ where for $m$ we
assign the value of $\frac{1}{2}$. Thus, as in \cite{Bar1}, the units we use for
$x$, $t$ and $p$ are $x=\frac{x_{cm}}{h}$, $t=\frac{t_{sec}}{mh}$ and $p=mv$.
That is, one may realize in this scaling that the velocities in $\frac{cm}{sec}$
are related to the mentioned parameters $x$ and $t$ by
$\frac{x_{cm}}{t_{cm}}=\frac{x}{mt}$. In order to  maintain the condition of $E>V$,
where $E$ is the energy of the passing wavepacket, we assign for the constant height
of the barriers the value of $V=2$.    \par
   The initial wavepacket which approach the multibarrier potential is expressed as 
  \begin{equation} \label{e7} \phi(t=0)=Re^2(P_{packet})+Im^2(P_{packet}), 
  \end{equation}   
  where $Re(P_{packet})$ and $Im(P_{packet})$ denote the real and imaginary 
 parts respectively of $P_{packet}$ from Eq (\ref{e6}).  The initial
wavepacket of Eq (\ref{e7}) is shown at the left hand side  Panel  
 of Figure 1. We note that by its definition the initial wavepacket $\phi(t=0)$ 
 from Eq (\ref{e7})   spreads with time without having to pass 
 through any potential (see
 Eq (\ref{e6})). 
 We are not interested here in this kind of known  spreading but, especially, 
 want to track and follow  the unknown chaotic-like  deformation of the packet
 due to its passage through the multibarrier potential. 
 Thus,   we numerically follow 
   the time evolution of the real and 
 imaginary parts 
 through the multibarrier  potential and obtain  the passing wavepacket
 as 
\begin{equation} \label{e8} \phi(t)=
 Re^2(P_{packet,V}) + Im^2(P_{packet,V}), \end{equation}   
 where $Re(P_{packet,V})$ and $Im(P_{packet,V})$ denote the real and  
 imaginary parts
 of $P_{packet}$ after passing through the multibarrier potential.  \par    
We note that the number  of the different chaotic wavepackets which  evolve from
the initial wavepacket of Eq (\ref{e7})  
is very large. For example, by slightly changing the ratio $c$ or by
adding or removing even one barrier results in a completely different wavepacket
 compared to the one which corresponds to the potential before the change. Since
 the passing wavepacket becomes chaotic  there is generally no rule that
 controls its form  or the correlation $C$ between 
 it and
 the initial wavepacket from Eq (\ref{e7}). As remarked, we show in the
 following section that for certain values of $N$ and $c$ one may predict the
 forms, and therefore the related correlation, of the passing wavepackets.  \par 
 The right hand side Panel of Figure 1 
 shows how the initial wavepacket from the left hand side expands and become
 deformed at time $t=6$ after passing through a ten-barrier potential whose ratio
 $c$ is unity. The influence of changing $N$   upon the passing wavepacket and  
 its corresponding
 correlation $C$ is further demonstrated at the left hand side Panel of
 Figure 2  which shows the waveform obtained for the same  $c$ and 
 $t$ as those of the right hand side Panel of Figure 1 but for a 15 barrier
 potential. Note that by increasing $N$ by 5 the wave packet becomes more
 chaotic and deformed compared to that at the right hand side of Figure 1. The
 dependence of the passing wavepacket upon $c$ is shown at the right hand side
 Panel of Figure 2 which is drawn for the same $N$ and $t$ as those of the
 left hand side but for $c=\frac{1}{9}$. That is, decreasing $c$ from unity to
 $\frac{1}{9}$ causes the passing wavepacket to become much less complex and
 chaotic compared to the form at the left hand side. \par
 
 \begin{figure}
\begin{minipage}{.49\linewidth}
\centering\epsfig{figure=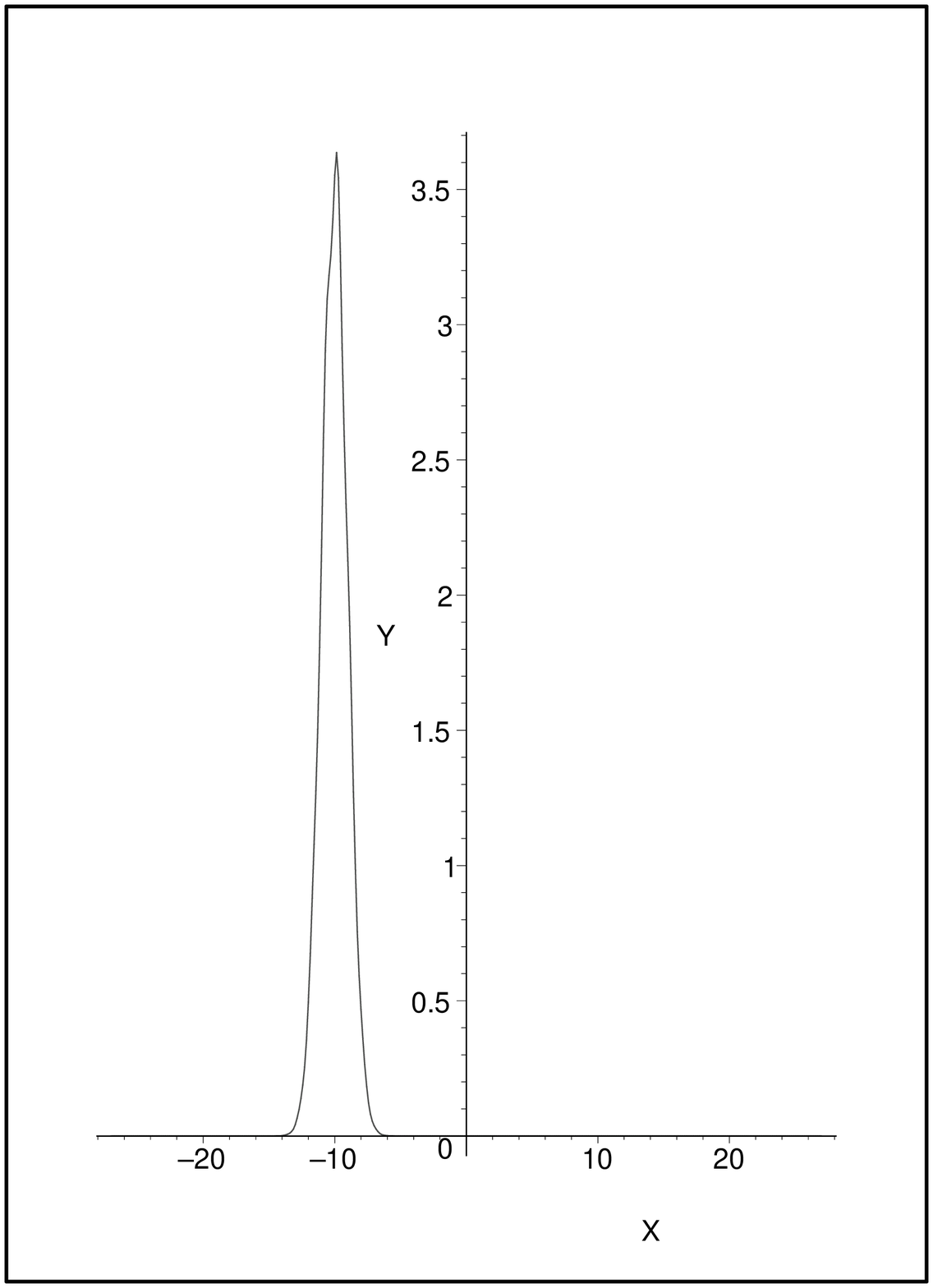,width=\linewidth}
  \end{minipage}  \hfill
  \begin{minipage}{.49\linewidth}
\centering\epsfig{figure=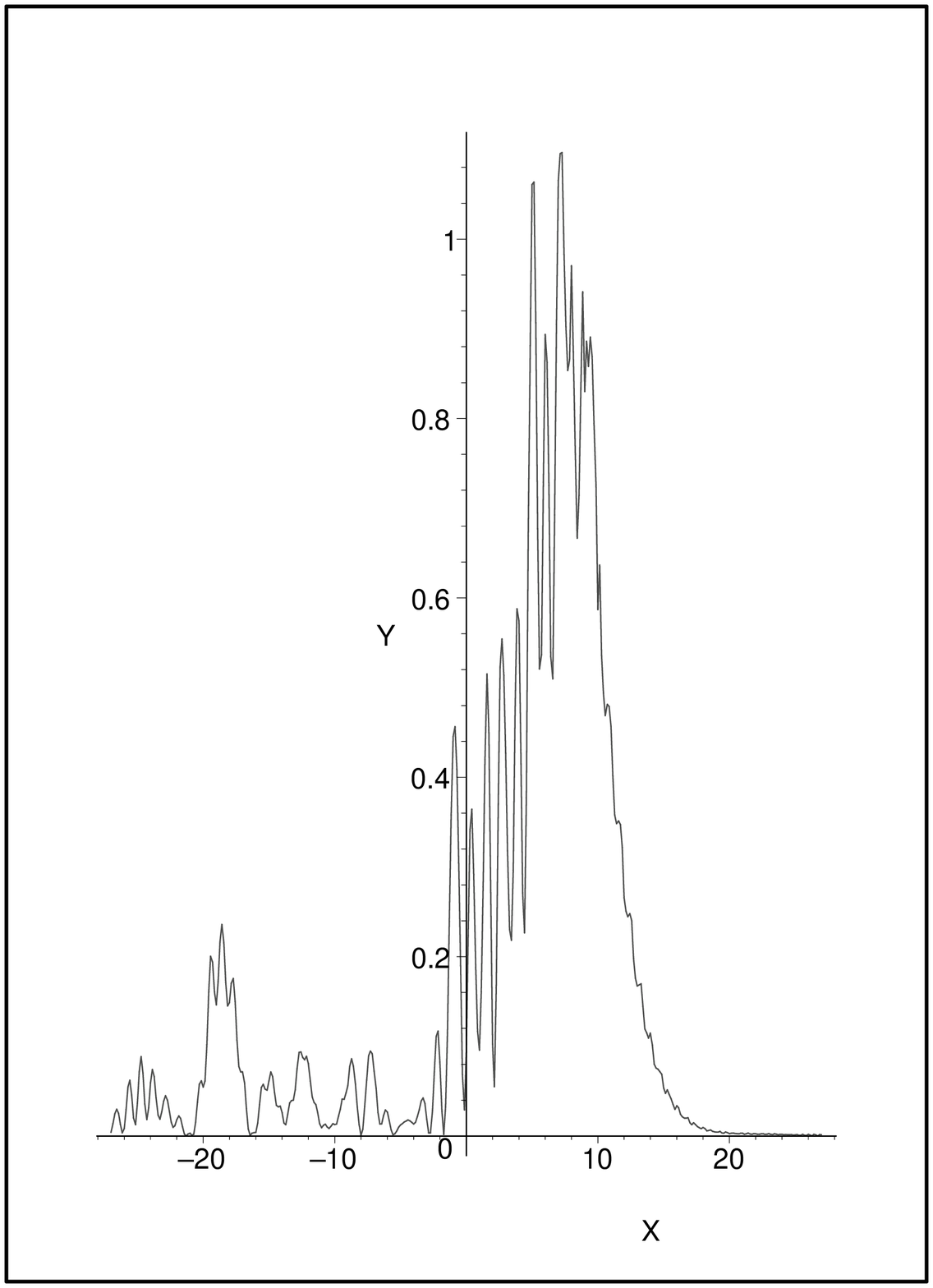,width=\linewidth}
\end{minipage}

\caption{The left-hand side figure shows the initial wavepacket from Eq
(\ref{e7}) as function of $x$ where $x$ is given in units of 
$\frac{x_{cm}}{h}$. The right hand side graph  shows this wavepacket 
 at time $t=6$ after passing 
a multibarrier potential composed of 10 barriers whose ratio of total interval to
total width is $c=1$.   Note how expanded and deformed the wavepacket becomes. 
Also note that the multibarrier potentials are not shown in this
figure and in Figure 2} 
\end{figure}

\begin{figure}
\begin{minipage}{.49\linewidth}
\centering\epsfig{figure=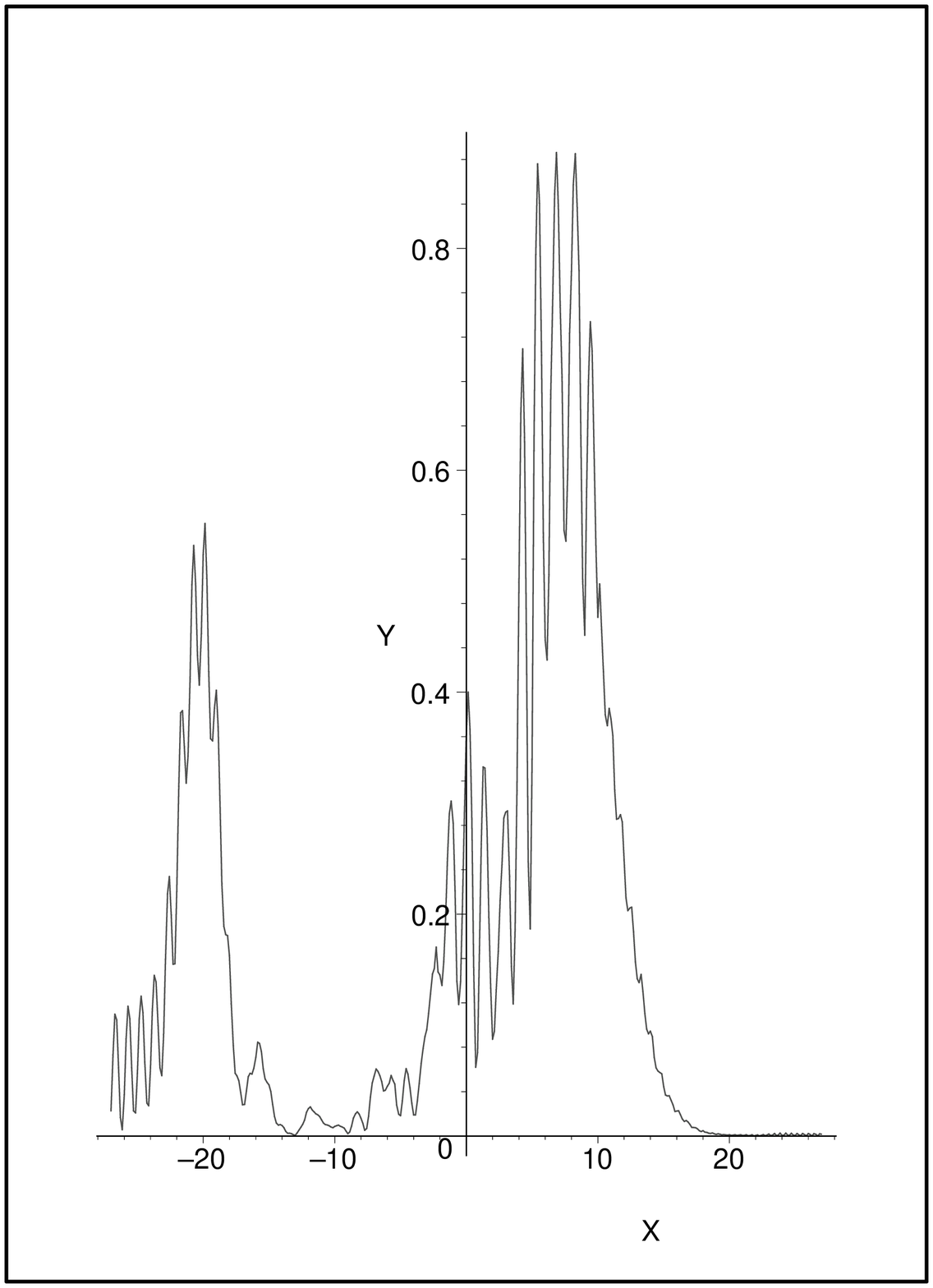,width=\linewidth}
\end{minipage}  \hfill 
\begin{minipage}{.49\linewidth}
\centering\epsfig{figure=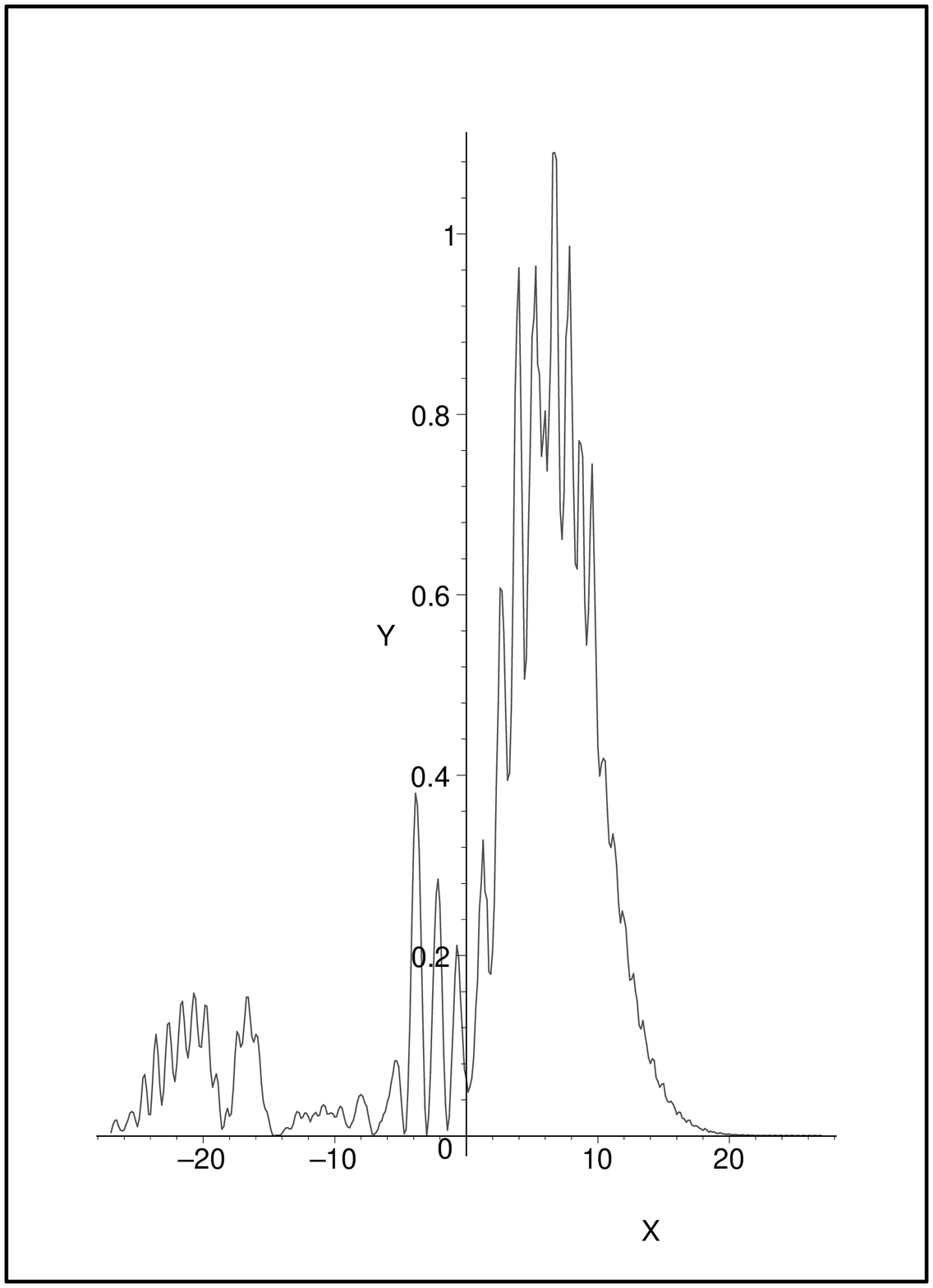,width=\linewidth}
\end{minipage}

\caption{The left hand side Panel shows how the initial wavepacket 
from Eq (\ref{e7})    changes  
  at time $t=6$ after passing  
a multibarrier potential composed of 15 barriers whose ratio 
 is $c=1$. Comparing this graph to  the right hand side 
Panel of Figure 1 one may
realize that  the wavepacket becomes more deformed and chaotic by increasing $N$
from 10 to 15 retaining the same ratio of $c=1$. The right hand side Panel  
shows  the   wavepacket obtained 
 at time $t=6$ for the same  15  barrier potential as that of the left hand side
 Panel but with a 
   ratio of $c=\frac{1}{9}$.  
   The decrease in $c$ results  in a wavepacket which is less complex 
  compared to that at the left hand side.}  
 \end{figure}

 We refer in the following to Table 1 which shows the
 correlation between the same initial wavepacket of Eq (\ref{e7}) and the
 passing one for 40  values of 
  $N$  and for six 
 different
 values of $c$: $c=4, \ \frac{7}{3},\ \frac{3}{2},\ 1, \ \frac{2}{3},\ 0.25$.  
 The spatial
 length of the multibarrier was fixed to $L=20$ and the time at which all 
 the passing
 wavepackets  were calculated is $t=6$ which corresponds to
 increasing 300 times  the mentioned time interval of $dt=\frac{1}{50}$. Thus, at
 this time the initial wavepacket have  passed through all the barriers 
 arranged along the fixed  length of $L$. Each of the tabulated values 
 of the correlation was numerically calculated using the Lanczos
 tridiagonalization method \cite{Lanczos,Cullum,Maple} which yields a tridiagonal matrix the values in its
 principal diagonal are the sought-for correlations. In this method the better
 and accurate result is given by the matrix element located at the bottom of
 the principal diagonal. Thus, the larger is the tridiagonal matrix the more
 accurate  becomes the correlation associated with this  matrix
 element. This is due to the large
 number of numerically running the  program which generally yields better
 results. We note, however, that the exact values of  the correlation is not of
 our main concern here   but especially we concentrate our attention upon
 its dependence on $c$ and $N$.  These  dependencies   may be established  
 even from a small tridiagonal matrix if we use consistently the same order of 
 it for all $c$ and $N$. Thus, we use for all the numerical work here a third 
 order Lanczos tridiagonalization matrix. An example of such a matrix is the
 following one which corresponds to $c=\frac{7}{3}$, $N=10$, $L=20$, $x_0=-10$, $p_0=3$, 
 $w_0=\frac{1}{2}$, $t=6$ and $V=2$. 
 $$ M(N=10,c=7/3)=\left[ \begin{array}{c c c} 2.630821735 & 33.07455189&0 \\ 33.07455189 & 
 1396.833502&52187.77697\\0&52187.77697&1669.69722 \end{array} \right]$$
 The values of the correlation between the passing wavepacket at the time $t=6$ 
 and the initial one  from Eq (\ref{e7}) are tabulated along the principal
 diagonal. The value of $1669.69722$ at the bottom of this diagonal is, as
 remarked,  more
 accurate than the two other values. 
 The tridiagonal matrix is symmetric which means that the off diagonal matrix
 elements  which  are symmetrically located  about the principal diagonal 
 are equal.  These off diagonal elements are the normalizing factors of the
 diagonal elements \cite{Cullum,Maple}. \par 
 Each one of the
  values tabulated in Table 1 is obtained from the bottom value of the
 principal diagonal of the corresponding tridiagonal matrix. The 40  
 rows in Table 1
 correspond to the three representative ranges of   $N=4,\ldots, 15$, 
 $N=31,\ldots, 40$ and $N=55,\ldots,72$
  and the six columns to the six values
 of the ratio $c=4, \ \ \frac{7}{3}, \ \ \frac{3}{2}, \ \ 1, \ \ 
 \frac{2}{3}, \ \ 0.25$.  As seen
 from the table the correlation ranges, for the specific values given here to the
 related parameters $L, \ w_0, \ x_0, \ p_0, \ V$  and $t$, over values which
 greatly differ among them. 
  Thus, in order to be able to
 graphically plot the correlation as function of $N$ (or $c$) one have to 
 scale
 the ordinate axis in a log basis as done in Figure 3-4. We must remark that
 Figures 3-5 are constructed not only from the tabulated values of Table 1 but
 also from other values which are not given in this table. That is, the
 correlation values used for Figures 3-5 are for all $N$ from $N=4+n, \ \ ,n=1,2,
 \ldots,68$  and for  $c=4, \ \ \frac{7}{3}, \ \ \frac{3}{2}, \ \ 1, \ \ 
 \frac{2}{3}, \ \ 0.25$.    From Table 1 and 
 Figures 3-4 one may realize that the correlation 
  changes in a stochastic and unexpected manner even when adding or
 removing only one barrier. Also, one may see that the larger values of the
 correlation $C$ are found at either large or small values of the ratio $c$ and
 the smaller values of $C$ are found at the intermediate values of $c$. This is
  shown in Figure 3 in which we compare  at the left hand side Panel 
 of it  the 
 correlation $C$ as function of $N$ for $c=4$ (continuous curve),  
  $c=\frac{7}{3}$ (dashed curve) and $c=\frac{3}{2}$ (dashdot curve). 
  At the right hand side Panel of Figure 3 we compare the
  correlation $C$, as function of $N$,   
  for $c=4$ (continuous curve),   $c=1$ (dashed curve) and 
 $c=\frac{2}{3}$ (dashdot curve).
  Remembering that
  the ordinate is scaled in a log basis one may realize, for example, 
   how large is the
  difference for $10 \le N \le 25$  between the correlation for $c=4$ and  
  those obtained for  
  $c=\frac{3}{2}$, $c=\frac{2}{3}$ and $c=1$.  
   Similar differences are demonstrated
 at the left hand side Panel  of Figure 4 where the correlation $C$, as
 function of $N$, for $c=0.25$   
  (continuous curve) is compared to those for  $c=\frac{7}{3}$ (dashed curve) 
 and $c=\frac{3}{2}$ (dashdot curve).  Note again the large differences 
 for $12 \le N \le 25$ between 
  the correlation $C$ obtained 
 for $c=\frac{3}{2}$ and those for $c=0.25$ and
 $c=\frac{7}{3}$.  The similarity of  the correlations for large and small
 $c$ is demonstrated at the right hand side Panel  of Figure 4 where we
 compare $C$ for $c=4$  (continuous curve) to that for $c=0.25$ (dashed curve).
Note, however,  the large difference between these correlations  for $28 \le
N \le 36$ where the $C$'s for  $c=4$ are much larger compared to those for
$c=0.25$. 
 \par 
 
\begin{figure}
\begin{minipage}{.49\linewidth}
\centering\epsfig{figure=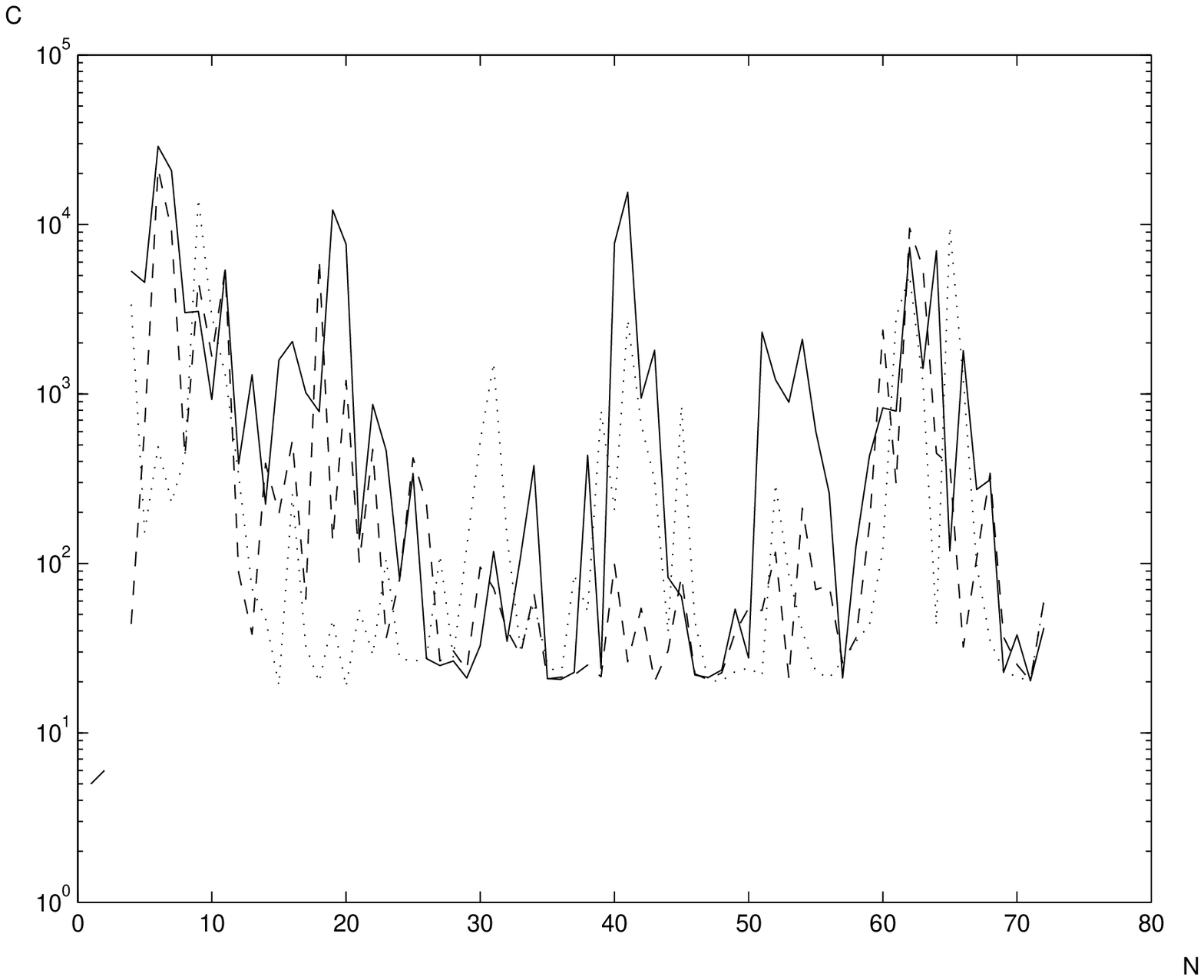,width=\linewidth}
  \end{minipage}  \hfill
\begin{minipage}{.49\linewidth}
\centering\epsfig{figure=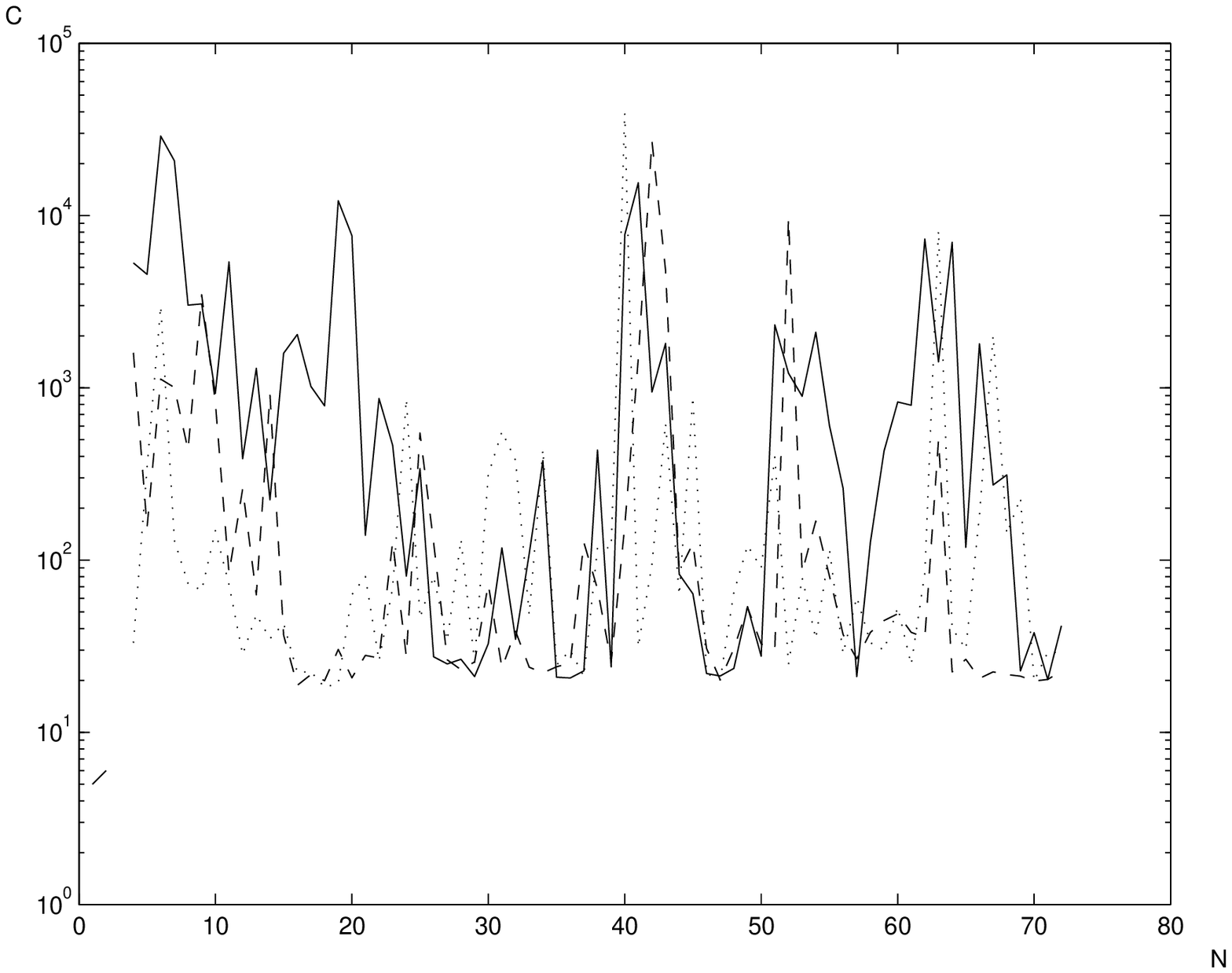,width=\linewidth}
\end{minipage}

\caption{At the left-hand side Panel we compare the correlation $C$ as
function of $N$ for $c=4$
(continuous curve) to those for $c=\frac{7}{3}$ (dashed curve) and
$c=\frac{3}{2}$
(dashdot curve). The right-hand side Panel compare the correlation  
$C$ for $c=4$
(continuous curve) to those for $c=1$ (dashed curve) and $c=\frac{2}{3}$
(dashdot curve). Remembering that the ordinate axis is scaled in a log basis one
may realize that the correlations for $c=4$, $c=\frac{7}{3}$, $c=\frac{3}{2}$, 
$c=\frac{2}{3}$ and $c=1$ widely differ from each other. 
 See, especially, the differences, for $10 \le N \le 25$,  
 among $c=4, \ \ c=\frac{3}{2}, \ \ c=\frac{2}{3}$ and $c=1$.   } 

\end{figure}

\begin{figure}
\begin{minipage}{.49\linewidth}
\centering\epsfig{figure=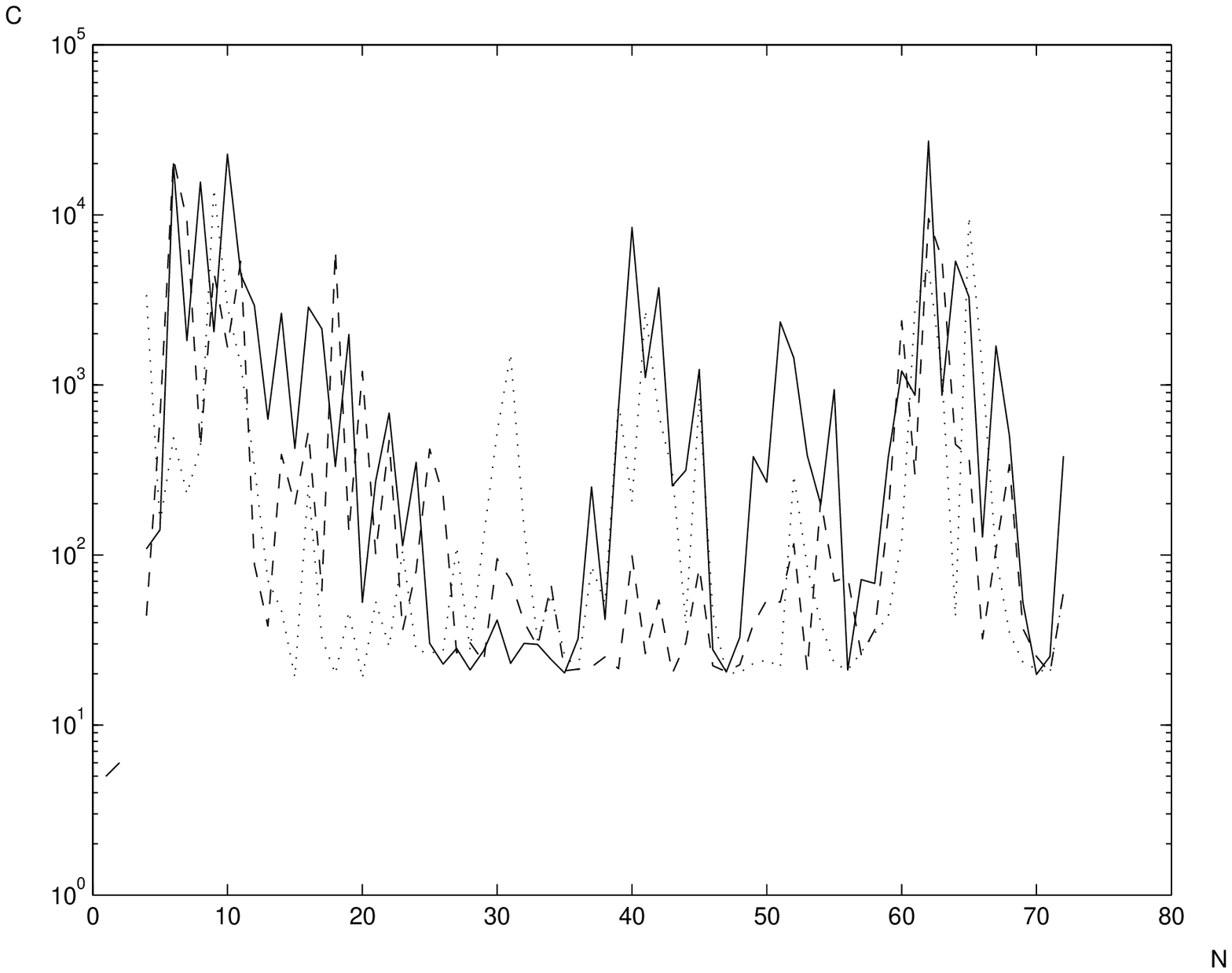,width=\linewidth}
  \end{minipage}  \hfill
\begin{minipage}{.49\linewidth}
\centering\epsfig{figure=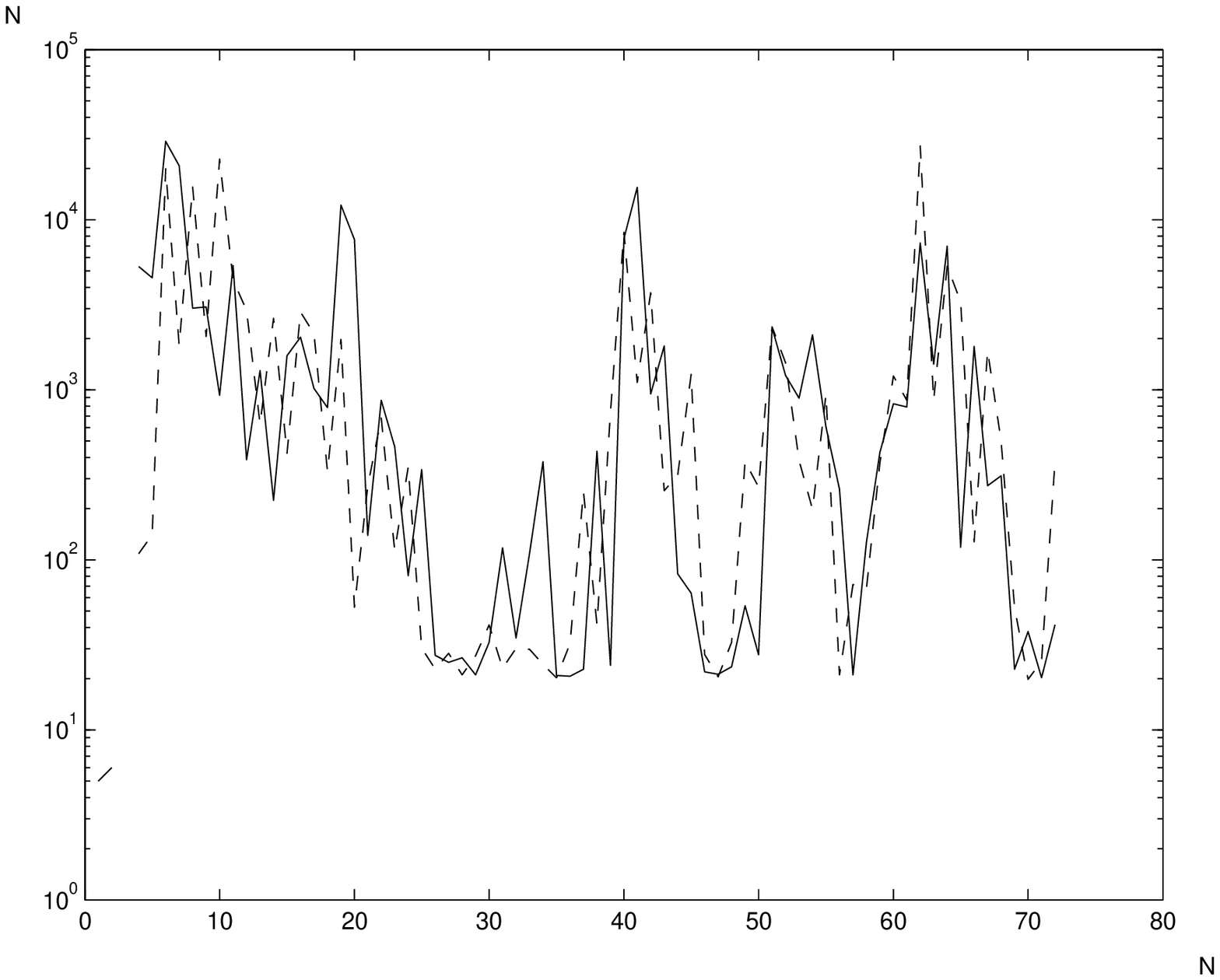,width=\linewidth}
\end{minipage}

\caption{At the left-hand side Panel we compare the correlation $C$ as
function of $N$ for $c=0.25$
(continuous curve) to those for $c=\frac{7}{3}$ (dashed curve) and
$c=\frac{3}{2}$
(dashdot curve). Taking into account that the ordinate axis is scaled in a log
basis one may realize, for example,   the large differences for $12 \le
N \le 25$  between the correlations 
$C$  for $c=\frac{3}{2}$ and those for $c=0.25$ and $c=\frac{7}{3}$. 
 At  the right-hand side Panel we compare the correlations  
$C$ for $c=4$
(continuous curve) to those for $c=0.25$ (dashed curve). From the similarity
between the two graphs (which, however, greatly differ for $28 \le N \le 36$)  
one may see that the
correlations are large for either large or small $c$.} 

\end{figure}

   From Table 1 and the
  corresponding Panels of Figures 3-4 one  finds the appropriate $N$  
  and $c$ for
  constructing a bounded one-dimensional multibarrier potential from which one
   may obtain  a large correlation between the initial and final wavepackets.

  \pagestyle{myheadings}
\markright{THE PERIODS OF CHAOS }
  \bigskip \noindent 
\protect \section{The periods of chaos  \label{sec3}}

As realized from the former section the wavepacket which passes through the
multibarrier potential becomes  deformed and chaotic  
and  the emerging   waveforms  are quite  different   
 even for neighbouring values of $N$ and $c$. 
    That is, 
the resulting  waveforms $\phi(t)$  and the corresponding correlations $C$ 
depend upon $c$ and
$N$ in such a manner that slightly changing either one of them  results in a
large change of $\phi(t)$ and $C$.  
We have, nevertheless, found that there exists  an unexpected order among the
multitude of the chaotic waveforms and the corresponding correlations. 
This order is reflected in 
periodicities which are clearly observed  for specific values of $N$ and $c$.  
That is, we
find that exactly the same identical wavepackets emerge from the potential
barriers when the number of the latter increases by specific numbers $P$ or by 
any  integral multiplication of them where the total spatial length of the
system  and the ratio $c$ remain fixed.  The specific periods $P$ are found 
to be of two kinds:
a large period of $P_L=140$ and a small one of $P_S=28$. That is, if the relevant $N$
(for a specific $c$) is periodic then exactly the same wavepacket emerge from
all the potentials which have $(N+nP)$  barriers where $n$ denote the whole
numbers $ 1, 2, \ldots$ and $P$ is either the large period of 140 or the
smaller one of 28. Note that since $5\cdot 28=140$ then any $N$ and $c$ which are
characterized as being periodic with the small period of 28 are also automatically       
 periodic with the larger period of 140. \par
 The large period of 140 is found to be very frequent and common for a large
 number of different $c$ and $N$ whereas the smaller period of 28 is rare. The criterion
 used here for characterizing any pair of $N$ and $c$ as periodic is that they
 have the same identical tridiagonal matrix for all $(N+nP)$ where $n=0, 1, 2,
 \ldots$ and $P$ is either 140 or 28. In Table 1 we have denoted  the
  multibarrier potentials which are periodic with the large 
   period of 140 by the word $p$ attached to the
 numerical values of the corresponding correlations $C$. 
 All the other values in Table 1 in which the word
 $p$ is absent are nonperiodic. Thus, as seen from the table there exists a
 large number of periodic multibarrier potentials which produce the same
 wavepackets and the same correlations when the number of barriers are increased
 by 140 or by any integral multiplication of it. It is found (see Table 1) that the
 smaller is the number of barriers $N$ the more frequent  is the number of
the  nonperiodic potentials and as  $N$ increases the periodicity of the
corresponding multibarrier potentials becomes more common and frequent. \par
In Figure 5 we have schematically drawn for the six values of $c$  
the correlations $C$ 
  as functions of $N$  from the point of view of whether they 
 are periodic with the large
 period of 140 or not. That is, each periodic $C$, which corresponds to some
 epecific  $N$ and $c$,   is denoted by a 
 point 
 and the absence of this point for some
 given   $N$ and $c$ signifies that the relevant $C$ is nonperiodic. Note that,
 as remarked, the values of $N$ used in this figure are not only those of 
 Table 1 but all the values of $N=4+n, \ \ n=0, 1, 2,\ldots,68$.  
  From the frequent occurence of the gaps in the horizontal lines of
 Figure 5 for small $N$ and  from the width of these gaps one may realize
 that   there exists a large number of nonperiodic
 correlations $C$ at these values of  $N$. The larger $N$ becomes the more 
 rare and narrow these gaps become which means that the number of 
  the periodic correlations $C$ increases for all values of $c$. 
  For very high values of $N$   (not shown in Figure 5) 
 the continuous linear sections  become very long for all $c$ which means that
 all the wavepackets as well as their corresponding correlations are periodic
 with the large period of 140.  \par  
Regarding  the smaller period of 28 we have found that it exists for the two 
pairs
of ($c=4$, $N=29$) and ($c=0.25$, $N=28$). That is, the same identical tridiagonal 
matrix, which implies the same emerging wavepacket and correlation, is obtained
for the $c=4$ case for all potentials  (arranged along the same fixed
length of $L=20$) which have  $(29+n\cdot 28)$  barriers where $n=0, \ 1,\ 2, \ \ldots$. 
Likewise, for
the $c=0.25$ case one obtains the same tridiagonal matrix, wavepacket and
correlation  (which are  not the same as  those  of the formerly discussed 
$c=4$ case) for all potentials which have $(28+n\cdot 28)$ barriers  
arrayed along the same fixed length of $L=20$ where  $n=0, \ 1,\ 2, \ \ldots$.\par
Another kind of ordered regularity which we have found among the multitude of all
the chaotic wavepackets is related to some specific tridiagonal matrices (and, 
therefore,  to their corresponding wavepackets and correlations) which remain
constant even  when  the ratio $c$ changes. That is, although
the general behaviour is the unexpected change and deformation 
 of the passing wavepacket
when  $c$  changes even slightly there exist,  nevertheless, specific values of
$N$ which are characterized as related to  correlations and  wavepackets 
 which 
retain their forms
even when $c$ is changed. For example, for $N=70$ we have found that the same
tridiagonal matrix (which means the same wavepacket and correlation) remains
constant for $1 \ge c \ge 0.25$. The same situation is also encountered for
$N=71$ where this time the constancy of the matrix, wavepacket and correlation  
are retained for the larger range of $4 \ge c \ge 1$.  The relevant tridiagonal
matrix for the last case is 
$$ M(N=71,1 \le c \le 4)= \left[ \begin{array}{c c c} 0.06426037493 &6.899198452&0 \\ 6.899198452& 
 68.29688359&562.8969638\\0&562.8969638&20.28518401 \end{array} \right]$$
 The same situation is again encountered for other values of $N$ 
 for which  one finds  constant 
 different matrices . These $N$'s and the corresponding
 ranges of $c$ over which the emerging wavepackets (and the appropriate
 correlations $C$) retain their forms are; at $N=47$  
  for $\frac{3}{2} \ge c \ge \frac{2}{3}$,  at $N=24$ 
  for $4 \ge c \ge \frac{7}{3}$ and   at $N=18$, $N=24$, $N=53$, and $N=123$ 
 for   
 $1 \ge c \ge \frac{2}{3}$. Note that all these values
 of $N$ are also  characterized as being periodic with the large period of 140.  
  Thus,
 one may realize that the constancy of the relevant tridiagonal matrices
 (and  the corresponding wavepackets and correlations) are retained not
 only for these specific  $N$ but also for all the other $N$'s obtained by
 increasing them  by 140 or by any integral multiplication of it.  
 In other words, this  behaviour of constant   waveforms (and correlations)
 when  $c$ changes is strongly related to the  previously mentioned 
  behaviour of periodic 
 waveforms (and correlations) when the number of barriers, for these specific
 $N$ and $c$,  increases by 140 or by
 any integral multiplication of it.

  \begin{figure}
\begin{minipage}{.70\linewidth}
\centering\epsfig{figure=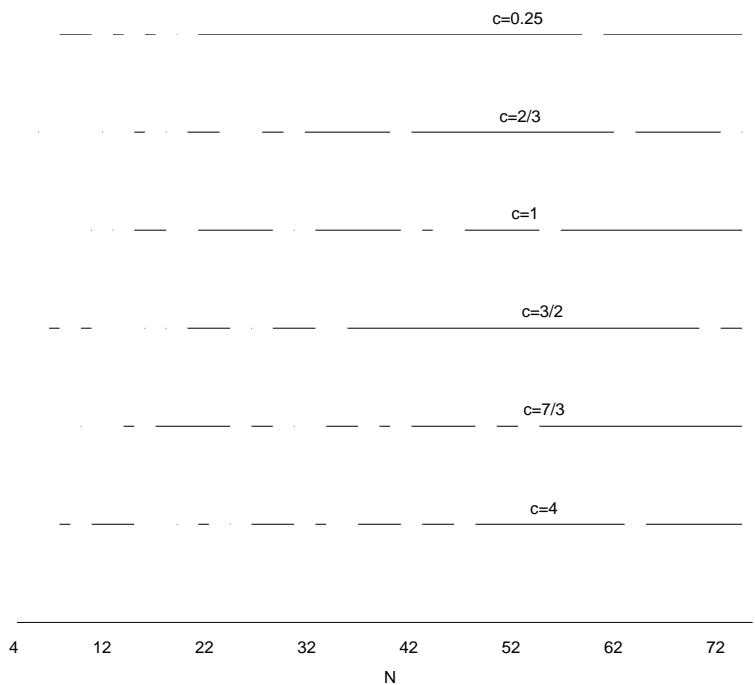,width=\linewidth}

   \caption{ From this figure  one may compare and find if  the correlations $C$, 
 for 
the six values of $c=0.25, \ \frac{2}{3},  \ 1, \ \frac{3}{2}, \
\frac{7}{3}, \ 4$,  have the large  period of $P=140$.
  The six values of $c$    label  the 6 horizontal lines. 
   A periodic $C$, with 
given $c$ and $N$,  is denoted by a 
point and the absence of this point signifies a nonperiodic $C$. Thus, as seen
from the figure the larger is $N$  in the abcissa the longer become the linear 
 continuous 
sections for all $c$ which denote that the number of periodic $C$ becomes 
large.  The wide gaps at these  lines for small $N$ signify that there
exists a large number of nonperiodic $C$ at these values of  $N$.} 

\end{minipage}
\end{figure}
  
  \pagestyle{myheadings}
\markright{CONCLUDING REMARKS}
  
  \bigskip \noindent

\protect \section{Concluding Remarks \label{sec4}}

We have discussed the chaotic deformed wavepackets which come out of a bounded
one-dimensional multibarrier potential and study the correlation of these 
wavepackets with the initial one. It has been shown, using the Lanczos
tridiagonalization method, that there exists an unexpected 
  order and regularity among 
the multitude of
all the possible chaotic wavepackets which come out of  this system. 
This order is characterized by the  existence of   two  periods 
through which one may obtain   the same wavepackets and correlations 
when
the number of barriers increase, for some specific $N$ and $c$, by either 
140 or 28  or by any integral multiplication of them.  
The  more common and frequent period is that of 140
whereas  the smaller one of 28 is rare. Any wavepacket and its
corresponding correlation which  is periodic with  the small period is 
also automatically periodic with the larger one. \par
 The
correlation $C$, as function of either $N$ or (and)  $c$,  between the passing 
wavepackets and the initial one is  stochastic and discontinuous as may
be realized from Table 1 and  Figures 3-4. 
One may see   from these figures and from Table
1  that the larger values of $C$ are obtained for $c \approx
4$ or $c \approx 0.25$. \par  
Another ordered behaviour  that we have found is related  to the constancy 
of the wavepackets and 
the corresponding correlations for specific  $N$ and for 
whole
ranges of $c$ that may be as large as $\Delta c=3$.  All these $N$'s are also
characterized as being periodic with the large period of 140. \par 
In summary, one may see that the chaos demonstrated by the bounded
one-dimensional multibarrier potential is an ordered and periodic phenomenon
especially for large $N$.

 \begin{table}
\caption{\label{table1} The table shows the correlations $C$ between 
the passing wavepacket at time $t=6$ and the initial
one from Eq (\ref{e7}).   
The rows correspond to the three ranges of  
 $ N =4+n, \ \ \  n=0, 1, \ldots 11$,  $ N =31+n, \ \ \  n=0, 1, \ldots 9$, 
 and $ N =55+n, \ \ \  n=0, 1, \ldots 17$. The columns correspond to 
the ratio $c$ 
for   $c=4,\  \frac{7}{3}, \ \frac{3}{2}, \  1, \ \frac{2}{3}, \ 0.25$. Any  
$N$ which have correlation $C$ with the word $p$ attached to it is periodic 
with 
the large period of 140. That is, any such  $N$ have exactly the same value of $C$ for all multibarrier potentials  
which have $(N+n\cdot 140)$ barriers where $n=1, 2, 3 \ldots$.}
     \begin{center}
      \begin{tabular}{|c|c|c|c|c|c|c|} 
        \  N&correlation  $C$  &correlations  $C$ &correlations $C$ &correlations  $C$ &
       correlations  $C$ &correlations  $C$ \\ & for \ \ c=4&for \ \
       c=$\frac{7}{3}$ & for \ \ c=$\frac{3}{2}$
       & for \ \ c=1 &for \ \ c=$\frac{2}{3}$ & for \ \ c=0.25 \\
   \hline \hline
$4  $&$  5.31202\cdot 10^3$&$ 4.392\cdot 10^1 $&$ 3.36365\cdot 10^3 $&$ 1.5939\cdot 10^3 $&$ 3.313\cdot 10^1 $&$ 1.0855\cdot 10^2$ \\
 $5  $&$4.55098\cdot 10^3 $&$ 6.380\cdot 10^2 $&$ 1.5324\cdot 10^2 $&$ 1.5478\cdot 10^2 $&$ 3.4617\cdot 10^2 $&$ 1.4041\cdot 10^2$ \\
$6  $ &$ 2.89038\cdot 10^4 $&$ 2.143528\cdot 10^4 $&$ 4.8818\cdot 10^2 $&$ 1.12423\cdot 10^3 $&$ 2.97818\cdot 10^3p $&$ 1.995463\cdot 10^4$ \\
$7$ &$ 2.076735\cdot 10^4 $&$ 9.20498\cdot 10^3 $&$ 2.2882\cdot 10^2p $&$ 1.00467\cdot 10^3 $&$ 1.3024\cdot 10^2 $&$ 1.82352\cdot 10^3$ \\
$8$ &$ 3.01606\cdot 10^3p $&$ 4.3025\cdot 10^2 $&$ 4.2984\cdot 10^2p $&$ 4.3489\cdot 10^2 $&$ 7.312\cdot 10^1 $&$1.555678\cdot 10^4p$ \\
$9$ &$ 3.06900\cdot 10^3p $&$ 4.59357\cdot 10^3  $&$1.421014\cdot 10^4  $&$3.47268\cdot 10^3 $&$ 6.664\cdot 10^1 $&$2.05691\cdot 10^3p$ \\
$10$ &$ 9.28730\cdot 10^2 $&$ 1.6697\cdot 10^3p $&$ 2.88651\cdot 10^3p $&$ 8.2338\cdot 10^2 $&$ 1.4825\cdot 10^2 $&$2.273285\cdot 10^4p$ \\
$11$ &$ 5.37380\cdot 10^3p $&$ 5.43254\cdot 10^3 $&$ 1.29778\cdot 10^3p $&$ 8.378\cdot 10^1p $&$ 7.195\cdot 10^1 $&$4.4193\cdot 10^3p$ \\
$12$ &$ 3.8843\cdot 10^2p $&$ 8.818\cdot 10^1 $&$ 3.316\cdot 10^2 $&$ 2.5926\cdot 10^2 $&$ 2.829\cdot 10^1p $&$ 2.93109\cdot 10^3$ \\
$13$ &$ 1.29587\cdot 10^3p $&$ 3.8130\cdot 10^1 $&$ 7.086\cdot 10^1 $&$ 6.284\cdot 10^1p $&$ 4.819\cdot 10^1 $&$6.2898\cdot 10^2p$ \\
$14$ &$ 2.2356\cdot 10^2p $&$ 3.903\cdot 10^2p  $&$4.743\cdot 10^1 $&$ 9.1362\cdot 10^2 $&$ 3.501\cdot 10^1 $&$2.63225\cdot 10^3p$ \\
$15$ &$ 1.58547\cdot 10^3p $&$ 1.9687\cdot 10^2p $&$ 1.905\cdot 10^1 $&$ 3.664\cdot 10^1p $&$ 4.278\cdot 10^1p $&$ 4.228\cdot 10^2$ \\
\hline
$31$ &$ 1.1767\cdot 10^2 $&$ 7.134\cdot 10^1 $&$ 1.52509\cdot 10^3p $&$ 2.368\cdot 10^1 $&$ 5.5717\cdot 10^2p $&$2.305\cdot 10^1p$ \\
$32$ &$ 3.4680\cdot 10^1p $&$ 4.026\cdot 10^1 $&$ 1.4336\cdot 10^2p $&$ 3.854\cdot 10^1p  $&$3.7924\cdot 10^2p $&$3.017\cdot 10^1p$ \\
$33$ &$ 1.08940\cdot 10^2p $&$ 2.887\cdot 10^1p $&$ 3.133\cdot 10^1 $&$ 2.39\cdot 10^1p $&$ 4.805\cdot 10^1p $&$2.981\cdot 10^1p$ \\
$34$ &$ 3.77230\cdot 10^2 $&$ 6.677\cdot 10^1p $&$ 5.853\cdot 10^1 $&$ 2.21\cdot 10^1p $&$ 4.3716\cdot 10^2p $&$2.427\cdot 10^1p$ \\
$35$ &$ 2.08800\cdot 10^1 $&$ 2.086\cdot 10^1p $&$ 2.562\cdot 10^1p $&$ 2.402\cdot 10^1p $&$ 2.46\cdot 10^1p $&$2.021\cdot 10^1p$ \\
$36$ &$ 2.06600\cdot 10^1p $&$ 2.128\cdot 10^1p $&$ 2.12\cdot 10^1p $&$ 2.522\cdot 10^1p $&$ 2.991\cdot 10^1p $&$3.214\cdot 10^1p$ \\
$37$ &$ 2.27400\cdot 10^1p $&$ 2.195\cdot 10^1 $&$ 8.573\cdot 10^1p$&$  1.2641\cdot 10^2p $&$ 2.11\cdot 10^1p $&$2.5054\cdot 10^2p$ \\
 $38$ &$4.34410\cdot 10^2p $&$ 2.518\cdot 10^1p $&$ 5.223\cdot 10^1p $&$ 6.78\cdot 10^1p $&$ 1.1804\cdot 10^2p $&$ 4.173\cdot 10^1p$ \\
 $39$ &$2.4010\cdot 10^1p $&$ 2.139\cdot 10^1p $&$ 7.9903\cdot 10^2p $&$ 2.556\cdot 10^1p $&$ 1.1795\cdot 10^2p $&$ 7.3825\cdot 10^2p$ \\
 $40$ &$7.75907\cdot 10^3p $&$ 9.886\cdot 10^1 $&$ 2.0648\cdot 10^2p $&$ 1.5902\cdot 10^2p $&$ 3.951835\cdot 10^4 $&$ 8.43398\cdot 10^3p$ \\
 \end{tabular} 
\end{center}
\end{table}

 \begin{table}
 \begin{center}
      \begin{tabular}{|c|c|c|c|c|c|c|}
      
     \  N&correlation $C$  &correlations $C$ &correlations  $C$
     &correlations  $C$ &
       correlations  $C$ &correlations  $C$ \\ & for \ \ c=4 &for \ \
       c=$\frac{7}{3}$ &for \ \ c=$\frac{3}{2}$ 
       & for \ \ c=1 &for \ \ c=$\frac{2}{3}$& for \ \ c=0.25 \\  
\hline \hline

$55$  &$6.0292\cdot 10^2p $&$ 7.002\cdot 10^1p $&$ 2.3000\cdot 10^1p $&$ 8.085\cdot 10^1p $&$ 1.153\cdot 10^2p $&$9.3828\cdot 10^2p$ \\
 $56$ &$2.6082\cdot 10^2p $&$ 7.428\cdot 10^1p $&$ 2.1090\cdot 10^1p $&$ 3.69\cdot 10^1p $&$ 2.865\cdot 10^1p $&$ 2.108\cdot 10^1p$ \\
 $57$ &$2.109\cdot 10^1p $&$ 2.591\cdot 10^1p $&$ 2.6780\cdot 10^1p $&$ 2.673\cdot 10^1p $&$ 6.079\cdot 10^1p $&$ 7.158\cdot 10^1p$ \\
$58$ &$ 1.2728\cdot 10^2p $&$ 3.629\cdot 10^1p $&$ 3.4680\cdot 10^1p $&$ 3.811\cdot 10^1p $&$ 3.308\cdot 10^1p $&$ 6.796\cdot 10^1$ \\
$59$ &$ 4.305\cdot 10^2p $&$ 1.6849\cdot 10^2p $&$ 4.3350\cdot 10^1p $&$ 4.462\cdot 10^1p $&$ 3.005\cdot 10^1p $&$3.7354\cdot 10^2p$ \\
$60$ &$ 8.2816\cdot 10^2p $&$ 2.38018\cdot 10^3p $&$ 1.2109\cdot 10^2p $&$ 4.874\cdot 10^1p $&$ 5.301\cdot 10^1p $&$1.20593\cdot 10^3p$ \\
 $61$ &$7.9283\cdot 10^2p $&$ 2.966\cdot 10^2p $&$ 2.84059\cdot 10^3p $&$ 3.805\cdot 10^1p $&$ 2.472\cdot 10^1 $&$ 8.6802\cdot 10^2p$ \\
 $62$ &$7.30202\cdot 10^3 $&$ 9.50662\cdot 10^3p $&$ 4.9941\cdot 10^3p $&$ 3.553\cdot 10^1p $&$ 8.411\cdot 10^1p $&$ 2.717475\cdot 10^4p$ \\
 $63$ &$1.41355\cdot 10^3p $&$ 5.51001\cdot 10^3p $&$ 1.15584\cdot 10^3p $&$ 4.9719\cdot 10^2p $&$ 8.17489\cdot 10^3p $&$ 8.6817\cdot 10^2p$ \\
 $64$&$ 6.98547\cdot 10^3p $&$ 4.4736\cdot 10^2p $&$ 4.3380\cdot 10^1p $&$ 2.241\cdot 10^1p $&$ 4.017\cdot 10^1p $&$ 5.33998\cdot 10^3p$ \\
 $65$ &$1.187\cdot 10^2p $&$ 3.811\cdot 10^2p $&$ 9.71896\cdot 10^3p $&$ 2.655\cdot 10^1p $&$ 3.04\cdot 10^1p $&$3.29676\cdot 10^3p$ \\
 $66$&$ 1.80058\cdot 10^3p $&$ 3.21\cdot 10^1p $&$ 1.22609\cdot 10^3p $&$ 2.057\cdot 10^1p $&$ 1.8292\cdot 10^2p $&$1.276\cdot 10^2p$ \\
 $67$ &$2.7297\cdot 10^2p $&$ 1.0732\cdot 10^2p $&$ 9.7360\cdot 10^1p $&$ 2.244\cdot 10^1p  $&$2.01296\cdot 10^3p $&$ 1.69351\cdot 10^3p$ \\
 $68$&$ 3.1167\cdot 10^2p $&$ 3.4067\cdot 10^2p $&$ 3.4780\cdot 10^1p $&$ 2.164\cdot 10^1p $&$ 1.4212\cdot 10^2p  $&$5.0016\cdot 10^2p$ \\
 $69$ &$2.274\cdot 10^1p $&$ 3.679\cdot 10^1p $&$ 2.3380\cdot 10^1 $&$ 2.12\cdot 10^1p $&$ 2.2945\cdot 10^2p $&$ 5.241\cdot 10^1p$ \\
 $70$ &$3.787\cdot 10^1p $&$ 2.53\cdot 10^1p $&$ 2.1300\cdot 10^1p  $&$1.983\cdot 10^1p $&$ 1.983\cdot 10^1p $&$ 1.983\cdot 10^1p$ \\
$71$ &$ 2.028\cdot 10^1p $&$ 2.028\cdot 10^1p $&$ 2.0280\cdot 10^1p $&$ 2.028\cdot 10^1p $&$ 2.912\cdot 10^1 $&$2.53\cdot 10^1p$ \\
 $72$ &$4.148\cdot 10^1p $&$ 5.996\cdot 10^1p $&$ 5.9960\cdot 10^1p $&$ 2.268\cdot 10^1p $&$ 2.905\cdot 10^1p $&$ 3.8058\cdot 10^2p$ \\
\end{tabular} 
\end{center}
\end{table}

\newpage

\pagestyle{myheadings}
\markright{REFERENCES}

\bigskip \bibliographystyle{plain}

\end{document}